\documentstyle[12pt]{article}
\def\ltsim{\raise 2pt \hbox {$<$} \kern-1.1em \lower 4pt \hbox {$\sim$}}
\def\gtsim{\raise 2pt \hbox {$>$} \kern-1.1em \lower 4pt \hbox {$\sim$}}
\begin{document}

\vspace{1.25cm}
\Large 

\def\etal{{\it et~al.~}}
\def\bsax{{\it Beppo-SAX~}}
\def\ginga{{\it Ginga~}}
\def\einstein{{\it Einstein~}}
\def\exosat{{\it EXOSAT~}}
\def\tenma{{\it Tenma~}}
\def\asca{{\it ASCA~}}
\def\rosat{{\it ROSAT~}}
\def\kbeta{$K_{\beta}$~}
\def\kalpha{$K_{\alpha}$~}
\def\ph{$~ph/cm^2~s~keV~$}

{\bf
%
\bsax OBSERVATION OF THE COMA CLUSTER
}
\normalsize

\vspace{0.5cm}

\newcommand{\au}[2]{#1$^{#2}$}
%
\au{R. Fusco-Femiano}{1}, \au{D. Dal Fiume}{2}, \au{L. Feretti}{3},
\au{G. Giovannini}{3,4}, \au{G. Matt}{5} and \au{S. Molendi}{6}

\vspace{0.5cm}

\newcommand{\ins}[2]{$^{#1}$ {\it #2}\\}
%
\ins{1}{Istituto di Astrofisica Spaziale, C.N.R., via del Fosso del Cavaliere, 
I-00133 Roma, Italy}
\ins{2}{TESRE, C.N.R., via Gobetti 101, I-40129 Bologna, Italy} 
\ins{3}{Istituto di Radioastronomia del CNR, Via P. Gobetti 101, I-40129
Bologna, Italy}
\ins{4}{Dipartimento di Fisica, Universit\'a di Bologna, via B. Pichat 6/2, 401
26 Bologna, Italy}
\ins{5}{Dip. di Fisica, Univ. ``Roma Tre", 
via della Vasca Navale 84, I-00146
Roma, Italy}
\ins{6}{Istituto di Fisica Cosmica, C.N.R., Via Bassini 15, I-20133 Milano, 
Italy}
\vspace{0.5cm}

ABSTRACT

%
We present first results of the BeppoSAX observation of the Coma 
Cluster. Thanks to the unprecedented sensitivity of the PDS instrument, 
the source has been detected up to $\sim$80 keV. There is clear evidence
for emission in excess to the thermal one above $\sim$25 keV, 
very likely of non-thermal origin. We have therefore, for the first
time, detected the long sought Inverse Compton emission on CMB photons
predicted in clusters, like Coma, with radio halos. Combining X and radio
observations, a value of 0.16 $\mu G$ for the volume-averaged 
intracluster magnetic field is derived.

\vspace{0.5cm}

INTRODUCTION

The Coma cluster
has been observed in December 1997, as \bsax AO-1 observation, for an 
exposure time of $\sim 91$ ksec 
with the Medium Energy
Concentrator Spectrometer (MECS) in the energy range 0.1-10 keV, and with the 
Phoswich Detection System (PDS) in the energy range 15-200 keV.
The main aim of this long observation 
was to search for
non-thermal hard X-ray radiation (HXR) exploting the unique 
capabilities of the PDS:
an overall sensitivity better than a few times $10^{-6}$\ph 
in the energy band 40-80 keV, a small field of view 
(FWHM=$1^{\circ}.3$, hexagonal) in order to reduce source confusion and
 wide energy coverage to measure the tail of the thermal emission
in the band 15-20 keV. 
Non-thermal hard X-ray radiation is predicted
in galaxy clusters showing a radio halo, such as Coma C present in the central
region of the Coma cluster (Willson 1970), due
to Compton
scattering of relativistic electrons by the Cosmic Microwave Background (CMB)
photons. The expected value of the spectral index of the non-thermal HXR 
is 1+$\alpha$, where $\alpha =1.34\pm 0.06$ is the slope of the radio spectrum
(Giovannini \etal 1993, ApJ, 406, 399). 
The combined radio and HXR detections would directly yield 
the opportunity to estimate the mean volume-averaged  
intracluster magnetic field, B, and the electron energy density, $\rho_e$.
The most recent attempt to detect hard X-ray radiation 
in the spectrum 
of the Coma cluster is due to the 
OSSE experiment, onboard the
{\it Compton GRO}. It reports upper limits that allow to get a lower limit 
on  B of
0.1$\mu$G and an upper bound of $3.5\times 10^{-13}~ergs~cm^{-3}$ on the
energy density of electrons with energies $\geq$500 MeV 
(Rephaeli, Ulmer $\&$ 
Gruber 1994). To get information on $\rho_e$ it is 
necessary to know the size
of the radio source and the distance of the cluster.
The importance of these 
determinations
is in that they are based essentially only on observables, whereas to
determine B and $\rho_e$ from the radio data alone recourse must be made to
additional theoretical assumptions, such as equipartion 
(Giovannini \etal 1993), or 
to polarization
studies of the Faraday rotation (Kim \etal 1990; Feretti \etal 
1995), but the results are model dependent. 

Throughout the paper
we assume a Hubble constant of $H_o = 50~km~s^{-1}~Mpc^{-1}$. The angular 
distance of $1^{\prime}$ corresponds to 40.6 kpc ($z_{Coma} = 0.0232$).

\vspace{0.5cm}

\begin{figure*}
\includegraphics{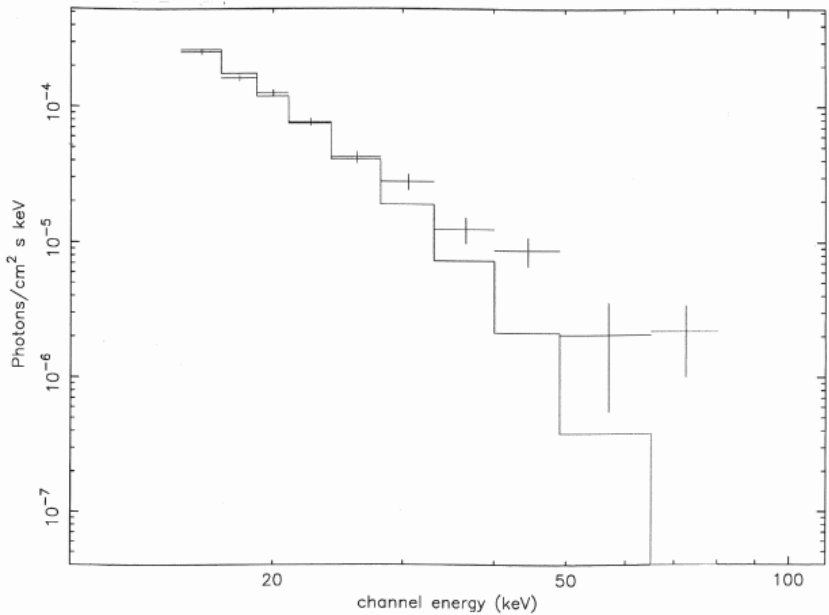}
\vspace{10 cm}
\caption{\it PDS data. The continuous line represents the best fit with a
thermal component at the average cluster gas temperature of 8.21 keV
(Hughes \etal 1993). Error bars are $1\sigma$}
\end{figure*}

DATA ANALYSIS AND RESULTS

The  observation with the MECS allowed 
to perform spatially resolved spectroscopy in the core of the
cluster with concentric annuli of $2^{\prime}$ within a region
of $8^{\prime}$ in radius ($\sim$325 kpc). The data analysis does not
show variations of the intracluster (IC) gas temperature and
metallicity with the radius. The average temperature results to be
$9.1\pm0.2$ keV and 0.257$\pm$0.026 the average iron abundance.
The temperature is somewhat larger than the typical value
obtained by larger field-of-view, non imaging instruments (see below),  
confirming the presence of a temperature gradient.
The \kbeta over \kalpha line ratio is not well determined:
\kbeta/\kalpha=$0.09^{+0.20}_{-0.09}$; the expected value at 9 keV being
$\sim$0.17 (Buote \etal 1998) for an optically thin plasma
(quoted confidence intervals are at $90\%$).
The MECS results will be presented in detail in a forthcoming paper.

The PDS detected a hard X-ray emission in the energy range 15-80 keV in the
Coma cluster spectrum, as shown
in Fig. 1. The continuous line represents the best-fit with a thermal component
at the average cluster gas temperature of 8.21($\pm$0.16) keV, as determined by 
the Large Area Counter (LAC) experiment onboard \ginga with a field of 
view of $1^{\circ}\times 2^{\circ}$ (Hughes \etal 1993), not 
much unlike 
that of the PDS ($1^{\circ}$.3, hexagonal). Different 
temperatures, in the range 7.5-8.5 keV, have been
reported by other X-ray observatories,  
but with a clear pattern: the larger
the field of view, the lower the measured temperature. 
With the PDS data alone it is not possible to perform complex 
spectral fits, while
MECS data are not of much help as
its field of view is much smaller than
that of the PDS. We therefore fixed the temperature of the thermal 
component to the accepted average
cluster value of 8.21 keV.  (But even adopting a temperature as high as 
9.1 keV, as observed by the MECS in the cluster core, the results are
substantially the same.)
The flux of the thermal
component is $\sim 3\times 10^{-10}~erg~cm^{-2}~s^{-1}$ in
the energy range 2-10 keV (the \ginga flux is 
$3.4\times 10^{-10}~erg~cm^{-2}~s^{-1}$ in the same energy range); the fit,
however, is very poor, with a
reduced $\chi^2$ of 3.2 for 9 d.o.f., due to a clear excess above
25 keV at a 4.5 $\sigma$ level.
The presence of a
second component in the PDS spectrum may be deduced also by the fit with a
single not fixed thermal component, which gives kT=10.71$\pm$0.76 with a
reduced $\chi^2$=1.98 for 8 d.o.f.:
The $\chi^2$ value has a significant
improvement by adding a second component, which may
be due to : i) a non-thermal component, or ii) a second
thermal component of temperature higher than 8.21 keV. 
Fitting the PDS data with
a thermal component at the average cluster gas temperature of 8.21 keV plus
a non-thermal component, the reduced $\chi^2$ value is 0.92 for 7 d.o.f. with
a large 1$\sigma$ confidence interval for the power law spectral index : 0.97-3.45; the flux of the non-thermal component is rather stable with respect to the
variation of the spectral index ($\sim 2\times 10^{-11}~erg~cm^{-2}~s^{-1}$
in the energy range 20-80 keV). As discussed in the Introduction, a non-thermal 
emission is predicted in clusters 
showing a radio halo, such as Coma C, due to inverse 
Compton scattering of relativistic electrons by the 3K 
microwave background photons. The theoretical value
of the spectral index is 2.34.
If we fix
the spectral index at the theoretical value of 2.34, we obtain an excellent
fit with a reduced $\chi^2$ of 0.83 for 8 d.o.f.; the flux of the thermal 
component (kT=8.21 keV) is $2.0\times 10^{-11}~erg~cm^{-2}~s^{-1}$ and
$2.2\times 10^{-11}~erg~cm^{-2}~s^{-1}$ for the non-thermal emission. 
On the other hand, 
if we consider a second thermal component, instead of the non-thermal component,
the fit requires a temperature greater than 40 keV. This unrealistic
value may be interpreted as a strong indication that the
detected hard excess is indeed due to a non-thermal mechanism. 

It is of interest to remark that the X-ray fluxes
measured by
\ginga and PDS coincide, within the errors, in their common energy range 
15-20 keV. The emission
detected by the PDS and \ginga LAC is essentially the tail of the thermal
radiation. Taking the theoretical value of 2.34 for the spectral index
of the hard Compton flux, the contribution of this component is not greater 
than 20$\%$ at 20 keV.
  
\vspace{0.5cm}

DISCUSSION AND CONCLUSIONS

The PDS onboard the \bsax satellite detected hard X-ray emission up to
energies of $\sim$80 keV in the spectrum of the Coma cluster.
The first check was to examine the possibility that the observed hard
emission is caused by contamination from a point source as an AGN. The 
sources present in the field of view of the PDS give a 
negligible contribution to the measured hard flux.
The PDS data can be fitted by fixing one thermal component to the average 
cluster gas
temperature of $\approx$8 keV, as determined by various X-ray experiments. 
The fluxes measured by the PDS and the  
\ginga LAC are in excellent agreement 
in their common energy range 15-20 keV. The field of view of the two 
experiments are approximately of the same order of size. In this energy range
the PDS and \ginga LAC measure essentially the tail of the 
thermal component.
As previously
discussed, the fits to the PDS data show an evident 
excess of emission which is hardly explainable by a second thermal component,
as the required temperature is unrealistically high
($>$40 keV), thus favouring a non-thermal mechanism.
The relativistic electrons, present in the IC medium of some clusters that
show extended regions of radio emission, have Compton scattering with the
CMB photons producing radiation that can reach MeV energies. 
Combining the syncrotron radio flux with the X-ray Compton flux 
it is possibile to derive, using only observables, the value of the 
mean volume-averaged IC magnetic
field, B, in the radio halo Coma C of size $\sim$1 Mpc, located in the central
region of the Coma cluster. This determination is independent of the radio
halo size and of the cluster distance. The radio data (Giovannini \etal 1993) 
 and   
the X-ray spectrum detected by the PDS implie a value
of B$\simeq 0.16~\mu G$. This value is not much different from the lower
limit of 0.1 $\mu G$ derived by the 2$\sigma$ upper bounds to the HXR 
reported by the
OSSE experiment.     
At
$\sim$50 keV, the HXR measurement of the PDS is lower by a factor $\sim$2
with respect to the upper limit. 
Knowing the radius size, R ($\simeq$1 Mpc), and the distance, 
d ($\simeq$138 Mpc), of the radio source we estimate 
an energy density of $6.8\times 10^{-14}~ ergs/cm^3$ for 
electrons with energies $\geq$ 500 MeV, using our derived value of B.
Methods to determine the IC magnetic field strength, B, are based on the 
detection
of energetic non-thermal photons, minimal arguments or measurements of Faraday
rotation of polarized radiation observed through the ICM. We have already 
cited the
lower limit of 0.1 $\mu G$ by the OSSE experiment. 
A greater lower limit of 0.4$ \mu G$ is derived from the upper limit 
to the gamma-ray
radiation above 100 MeV, due to relativistic bremsstrahlung, obtained by the  
EGRET experiment (Sreekumar \etal 1996), assuming an 
extrapolation of the radio 
spectrum to lower frequencies. This model requires the presence of
energetic cosmic rays in the cluster that interact with the hot ICM. Using
minimal energy arguments, Giovannini \etal (1993) derive 
an equipartition magnetic field of  
$\sim 0.4h_{50}^{2/7}~ \mu G$. Greater estimates of B are given by Faraday 
rotation
measurements of the polarized radiation seen through the hot ICM of the Coma
cluster, whose gas density profile is derived by X-ray observations. These
values of B result 1.7$h_{50}^{1/2}~\mu G$ (Kim \etal 1990) and 
$\sim 6h_{50}^{1/2}~\mu G$ (Feretti \etal 1995), although the last authors
suggest also the presence of a weaker magnetic fiel component 
($\sim 0.2h_{50}^{1/2}~\mu G$), uniform on the cluster core radius scale.
The value of the mean volume-averaged IC magnetic field, derived using the 
detection of non-thermal HXR by the PDS 
(B$\simeq 0.16~\mu G$),  
is below the estimates resulting from the Faraday rotation measurements. This
reflects the fact that the PDS observation measures the magnetic content of
a large region of the Coma cluster. Besides, the value of B 
derived by the Faraday rotation measurements may be altered by the  
presence of local density
peaks in a clumpy IC.  
Very attractive suggestions have been proposed regarding the origin of the
IC magnetic field, as that in which galaxy motion may drive a turbulent 
dynamo in order
to amplify faint seed fields to $\mu G$ fields. However, the origin of
the diffuse IC magnetic field in clusters of galaxies has not yet been 
clarified. 
(Ruzmaikin \etal 1989; De Young 1992). 
 
\underline{Acknowledgments} 

We thank P.Grandi and L.Piro for useful suggestions regarding the data
analysis.

\end{document}